\documentclass[journal]{IEEEtran} 

\usepackage{amsmath}
\DeclareMathAlphabet{\orgmathcal}{OMS}{cmsy}{m}{n}
\usepackage{newtxmath} 
\usepackage{bm}  
\usepackage{graphicx}
\graphicspath{{./}}
\usepackage{datetime}
\usepackage{caption}
\usepackage{subcaption}
\usepackage[percent]{overpic}
\usepackage{placeins} 
\usepackage{enumitem}
\usepackage{outlines}
\usepackage{enumitem}
\usepackage[ruled,vlined]{algorithm2e}
\usepackage{tabularx,colortbl}
\usepackage{makecell}
\usepackage{paralist}
\usepackage{algorithmicx }
\usepackage{color}
\usepackage{cite}
\usepackage{cleveref}
\usepackage[percent]{overpic}
\Crefname{figure}{Figure}{Figures}  
\crefname{figure}{Fig.}{Figs.}  
\usepackage{xargs}                     
\usepackage[pdftex,dvipsnames]{xcolor}

\newcommand{\NN}{N}

\begin{document}

\title{Single Snapshot Super-Resolution DOA Estimation for Arbitrary Array Geometries}

\author{Anupama~Govinda Raj,~\IEEEmembership{Student Member,~IEEE,}
	and James~H.~McClellan,~\IEEEmembership{Life~Fellow,~IEEE}
\thanks{This work was supported by the Franklin Foundation, John \& Marilu McCarty Chair.}
\thanks{Authors are with the School
	of Electrical and Computer Engineering, Georgia Institute of Technology, Atlanta,
	GA, 30332 USA (e-mail: agr6@gatech.edu; jim.mcclellan@ece.gatech.edu).}}
\markboth{IEEE Signal Processing Letters}{}

\IEEEpubid{Copyright \copyright~2018 IEEE}
\maketitle

\begin{abstract}

We address the problem of search-free direction of arrival (DOA) estimation
for sensor arrays of arbitrary geometry under the challenging
conditions of a single snapshot and coherent sources. 
We extend a method of search-free super-resolution beamforming, originally applicable only for uniform linear arrays, to arrays of arbitrary geometry. 
The infinite dimensional primal atomic norm minimization problem in continuous angle domain is converted to a dual problem.
By exploiting periodicity, the dual function is then represented with a trigonometric polynomial using a truncated Fourier series.
A linear rule of thumb is derived for selecting the minimum number of Fourier coefficients required for accurate polynomial representation, based on the distance of the farthest sensor from a reference point.
The dual problem is then expressed as a semidefinite program 
and solved efficiently.
Finally, the search-free DOA estimates are obtained through polynomial rooting, and source amplitudes are recovered through least squares. 
Simulations using circular and random planar arrays show perfect DOA estimation in noise-free cases.

\end{abstract}


\section{Introduction}\label{sec:intro}
\IEEEPARstart{E}{stimating}
the direction of arrival (DOA) of sources using an array of sensors is an important problem having numerous practical applications 
\cite{skolnik_radarsystems_1980,burdic_underwater_1980,godara_commn_1997,vanTrees_optimumArrayProc_2004,roh_mmavebeamform_5g_2014}. 
DOA estimation can be especially challenging when sources are coherent and snapshots are limited
as happens with multipath propagation and fast moving sources.
{Under these conditions, high resolution adaptive DOA estimation approaches such as 
MVDR \cite{capon_mvdr_1969}, MUSIC \cite{schmidt_music_1986} and covariance matching  
methods
\cite{stoica_spice_likes_2012,ottersten_stoica_covariancematching_1998}
fail due to self signal cancellation and\,/\,or inaccurate estimation of the spatial covariance matrix.}

Compressed sensing (CS) \cite{donoho_cs_2006,candes_cs_2006} and sparsity based approaches have been applied to the 
DOA estimation problem \cite{gurbuz_cs_beamform_2008,xenaki_compressive_beamforming_2014 } which can tackle coherent sources and single snapshots.
However, CS approaches suffer from the \textit{off-grid} problem 
\cite{chi_basis_mismatch_2011} when the actual sources do not fall on the discrete grid of angles used to form the basis. 
To improve performance, greedy algorithms with a highly coherent dictionary (finer search grids) are used
in \cite{duarte_baraniuk_spectral_cs_2013,fannjiang_coherence_2012},
but they are computationally demanding.
The off-grid DOA approaches \cite{zhu_perturbedCS_2011,yang_robust_perturbedCS_2012,austin_dynamicdictionary_2013,dai_FDD_MIMOchest_2018} applicable for arbitrary arrays use a Taylor series approximation of array steering vectors on fixed grids, or iterative methods with dynamic grids to tackle the grid mismatch.
{However, their performance depends on the approximation accuracy 
or they involve nonconvex optimization, providing only local convergence.}
Recent \textit{super-resolution} methods based on convex optimization
\cite{tang_cs_offgrid_2013,candes_math_theory_super_resoln_2014, candes_super_resoln_noisy_2013,xenaki_gridfree_2015}, 
in which the basis is formed in the continuous angle domain, i.e., no grid, eliminate
the off-grid problem and provide highly accurate solutions, but they are not applicable to arbitrary array geometries.

In this letter, we develop a search-free DOA estimation method for arrays of arbitrary geometry under the challenging conditions of coherent sources and a single snapshot. 
This extends super-resolution DOA estimation to arrays of arbitrary geometry. 
First, we express the DOA estimation problem for arbitrary geometry as an \textit{atomic norm}\cite{chandrasekaran_atomicnorm_2012} minimization problem in the continuous angle domain, 
which is then solved as a dual maximization problem.  
By exploiting the periodicity and band-limited nature of the dual function, we can approximate it with a \emph{finite trigonometric polynomial} using Fourier series (FS).
The proposed approach is motivated by \cite{rubsamen_gershman_FDrootmusic_2009},
in which root-MUSIC is extended to arbitrary arrays using a Fourier series based approximation.
Our approach is also related to the manifold separation technique for root-MUSIC in \cite{doron_wavefield_1994,doron_wavefield_algos_1994,belloni_MS_arbarrays_2007}, 
when the DFT is used to compute the manifold approximation matrix for arbitrary arrays.
The modified dual problem can then be expressed as a \emph{finite} semidefinite program (SDP), and solved efficiently.
Finally, the search-free DOA estimates are obtained through polynomial rooting of a nonnegative polynomial formed from the dual polynomial.
\vspace*{-0.1mm}
\section{Data Model} 
Consider an arbitrary geometry array of $ M $ sensors, 
which receives signals from $ L $ narrowband far-field sources with complex amplitude $ s_l $ and azimuth DOA $ \theta_l $, 
$ l = 1, \dots, L $.
We define the function $ x(\theta) $ in the continuous angle domain $\theta\in(-\pi,\pi]$ with impulses for the $L$ sparse sources \cite{xenaki_gridfree_2015} as
\begin{equation}
  x(\theta) =  \sum_{l=1}^{L} s_l \delta(\theta- \theta_l).
  \label{eq:xthetaDefined}
\end{equation} 
We express the observed array snapshot vector $\bm{y} \in \mathbb{C}^{M} $ as 
\begin{equation}
\label{eq:MeasurementModel}
\bm{y} = \mathcal{S} x,  \;\;\;\text{where}\; y_m = \int\limits_{-\pi}^{\pi} a_m(\theta) x(\theta) d\theta, \;\;\;\; m=1,\ldots,M.
\end{equation} 
The linear measurement operator $ \mathcal{S} $ represents the array manifold surface over $\theta$, whose $ m $-th component $a_m(\theta)$  
is the response of the $ m $-th sensor for a source at direction $\theta$.
\begin{equation}
  a_m(\theta) = e^{-j2\pi f \tau_m(\theta)},
  \label{eq:ar-spDefined}
\end{equation}
where $ \tau_m(\theta)$ is the propagation delay with respect to a reference.%
\footnote{We prefer to study $a_m(\theta)$ as a function of $\theta$. On the other hand, at a specific angle $\theta_1$, $[a_m(\theta_1)]\in\mathbb{C}^{M}$ is the steering vector for direction $\theta_1$.}
For narrowband sources of frequency $f$ and propagation speed $v$, the wavelength is $\lambda = v/f$. 
Using $ \tau_m(\theta) = {\langle\bm{p}_m, \bm{u}_{\theta}\rangle}/{v}$, we simplify the exponent in \eqref{eq:ar-spDefined} as 
\begin{equation} 
    2\pi f \tau_m(\theta) 
  = 2\pi (|\bm{p}_m|/\lambda) \cos(\theta - \angle \bm{p}_m),
  \label{eq:ar-spExponent}
\end{equation} 
where $ \bm{p}_m $ is the position vector of the $ m $-th sensor with respect to a reference, and $\bm{u}_{\theta}$ is a unit vector in source direction $\theta$.

\section{Super-resolution DOA for Arbitrary Arrays}\label{sec:offgridDOA} 
A brief review of the noise-free primal DOA estimation problem and its dual is given here, before introducing the Fourier series representation of the dual function.
Assuming the sources are sparse in angle, $x(\theta)$ in \eqref{eq:xthetaDefined} could be recovered \cite{candes_math_theory_super_resoln_2014,xenaki_gridfree_2015} via
\begin{equation}\label{eq:gridfree_primal}
\min_x \, \lVert x \rVert^{\phantom{.}}_{\orgmathcal{A}}, \quad {\rm s.t.} \quad \bm{y} = \mathcal{S} x,
\end{equation}
where $ \lVert . \rVert^{\phantom{.}}_{\orgmathcal{A}} $ denotes the {atomic norm} \cite{chandrasekaran_atomicnorm_2012} which is a continuous analogue of the $ l_1 $ norm, i.e., \(
\lVert x \rVert^{\phantom{.}}_{\orgmathcal{A}} = \sum_{l=1}^{L} \lvert s_l \rvert.
\)
Here $ \mathcal{S}  $ does not represent Fourier measurements, unlike \cite{tang_cs_offgrid_2013,candes_math_theory_super_resoln_2014,candes_super_resoln_noisy_2013,xenaki_gridfree_2015}.

The primal problem \eqref{eq:gridfree_primal} is infinite dimensional and difficult to solve, so we work with the dual problem.
Using the Lagrangian and the property that the \textit{dual function} defined by
$ \mathcal{S}(\theta)^H \bm{c} $ has unit magnitude in the direction of actual sources, irrespective of geometry, the primal problem can be converted to the following dual maximization problem with dual variable $\bm{c}$ (see details in \cite{candes_math_theory_super_resoln_2014 }, and also in \cite{xenaki_gridfree_2015, tang_cs_offgrid_2013})
\begin{equation}\label{eq:gridfree_dual}
\max_{\bm{c}\in \mathbb{C}^M } \operatorname{\Re}\{\bm{c}^H \bm{y}\},\quad \text{s.t.\ \;}  \lVert \mathcal{S}(\theta)^H \bm{c} \rVert _{\infty} \le 1.
\end{equation}
For a uniform linear array (ULA), the dual function $ \mathcal{S}(\theta)^H \bm{c} $ is, in fact, an \mbox{$(M-1)^\text{th}$} degree polynomial in $e^{j\theta}$, and \eqref{eq:gridfree_dual} is then solved using an SDP \cite{candes_math_theory_super_resoln_2014, candes_super_resoln_noisy_2013,xenaki_gridfree_2015}. 
The polynomial structure arises from the fact that sensor delays in 
(\ref{eq:ar-spDefined},\,\ref{eq:ar-spExponent}) for a ULA are integer multiples of a constant.
For arbitrary arrays, $ \mathcal{S}(\theta)^H \bm{c} $ cannot be directly expressed as a polynomial, 
but we overcome this difficulty with a Fourier domain (FD) representation of the dual function that provides a polynomial form for the SDP. 

\vspace*{-2.3mm}
\subsection{Fourier Domain Representation of the Dual Function}
The function $ b(\theta) = \mathcal{S}(\theta)^H \bm{c}$ is a linear combination of smooth (band-limited) periodic functions, $a^*_m(\theta)$, 
so it is also periodic in $ \theta $ with period $ 2\pi $. 
Thus, $ b(\theta) $ has a Fourier series (FS)
which can be truncated if its Fourier coefficients $ {B}_{k}\!\approx\!0 $ for $|k|>N$.
In fact, each $a^*_m(\theta)$, being periodic, has a FS with coefficients $\alpha_m[k]$, from which we construct the Fourier coefficients $B_k$ via $B_k = \sum_m \alpha_m[k] c_m$.
Thus, we have
\begin{equation}
 b(\theta) = \mathcal{S}(\theta)^H \bm{c}
  = \sum_{k=-N}^{N} \sum_{m=1}^{M}(\alpha_m[k] c_m) e^{jk\theta},
  \label{eq:BkFromAlphak}
\end{equation}
which is a \emph{finite degree} polynomial in $ z = e^{j\theta} $.
As a result, we can determine $N$ for FS truncation by examining the FS coefficients at each sensor, $\alpha^{\phantom{.}}_m[k]$, which depend solely on the array geometry and not on the measured signals. 
{Thus, \emph{the Fourier representation can be done off-line.}}
 
The Fourier coefficients $\alpha^{\phantom{.}}_m[k]$ {can be computed with high accuracy from samples of $a^*_m(\theta)$ using the DFT,
assuming a sufficiently large number of DFT points $(P=2N+1)$ for dense sampling in $\theta$ that eliminates any spectral aliasing \cite{rubsamen_gershman_FDrootmusic_2009,DSP_First_2015}.}
\begin{equation}
 \hat\alpha^{\phantom{.}}_m[k] \simeq
    (1/P)\sum_{l=-N}^{N} a^*_m(l\Delta\theta) e^{-j(2\pi/P) l k},
   \label{eq:P-ptDFT}
\end{equation}
where $\Delta\theta=2\pi/P$, and $k=-N,\ldots,0,1,\ldots,N$.
Note that circular indexing of the DFT is exploited in \eqref{eq:P-ptDFT}.

\begin{figure}[htbp]
	\begin{subfigure}{.5\columnwidth}
		\centering
		\begin{overpic}[width=1\textwidth]{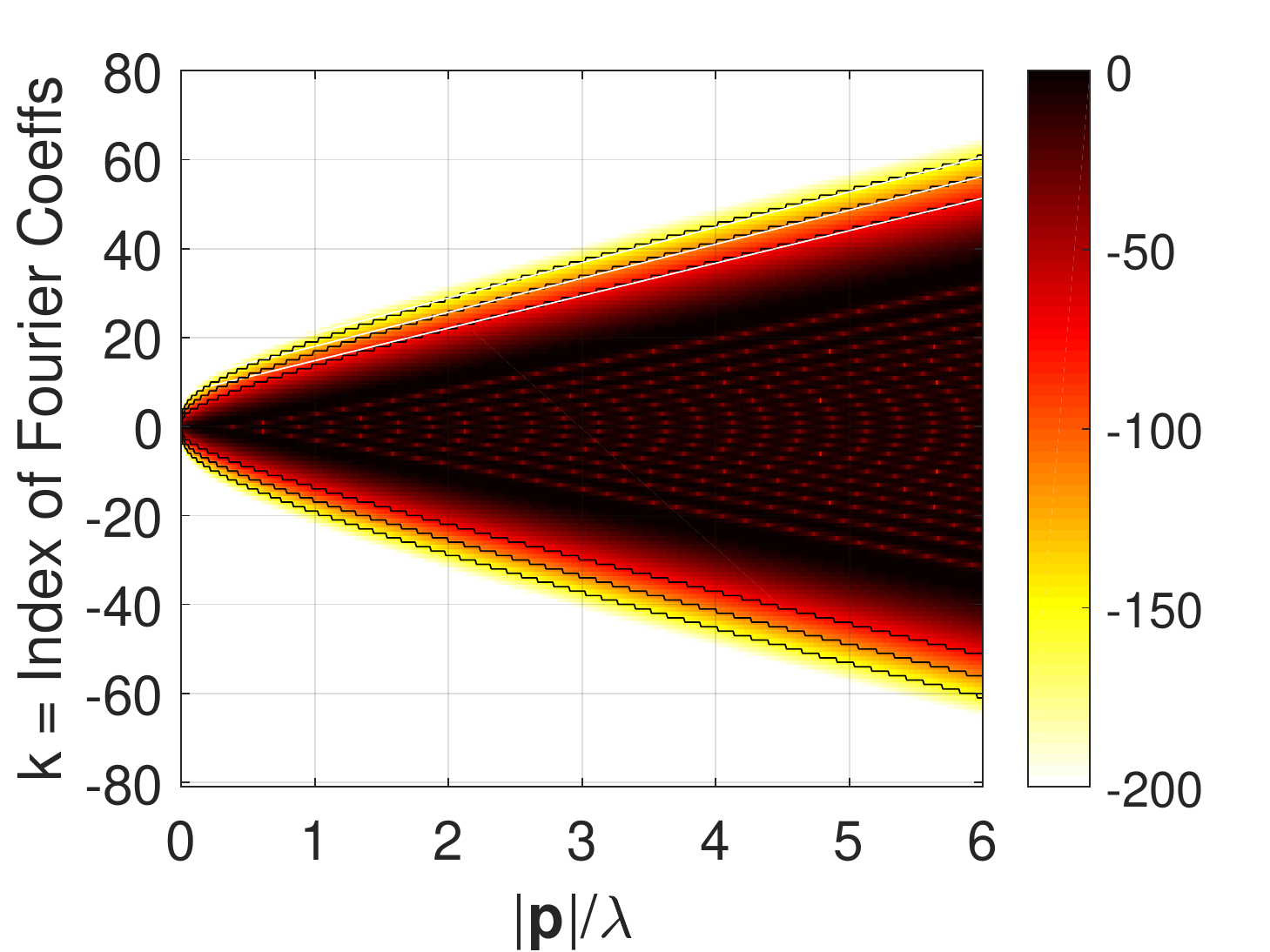}
			\put (17,63 ) {{\small {(\textbf{a})}}}
		\end{overpic}
	\end{subfigure}
	\begin{subfigure}{.49\columnwidth}
		\centering
		\begin{overpic}[width=1\textwidth]{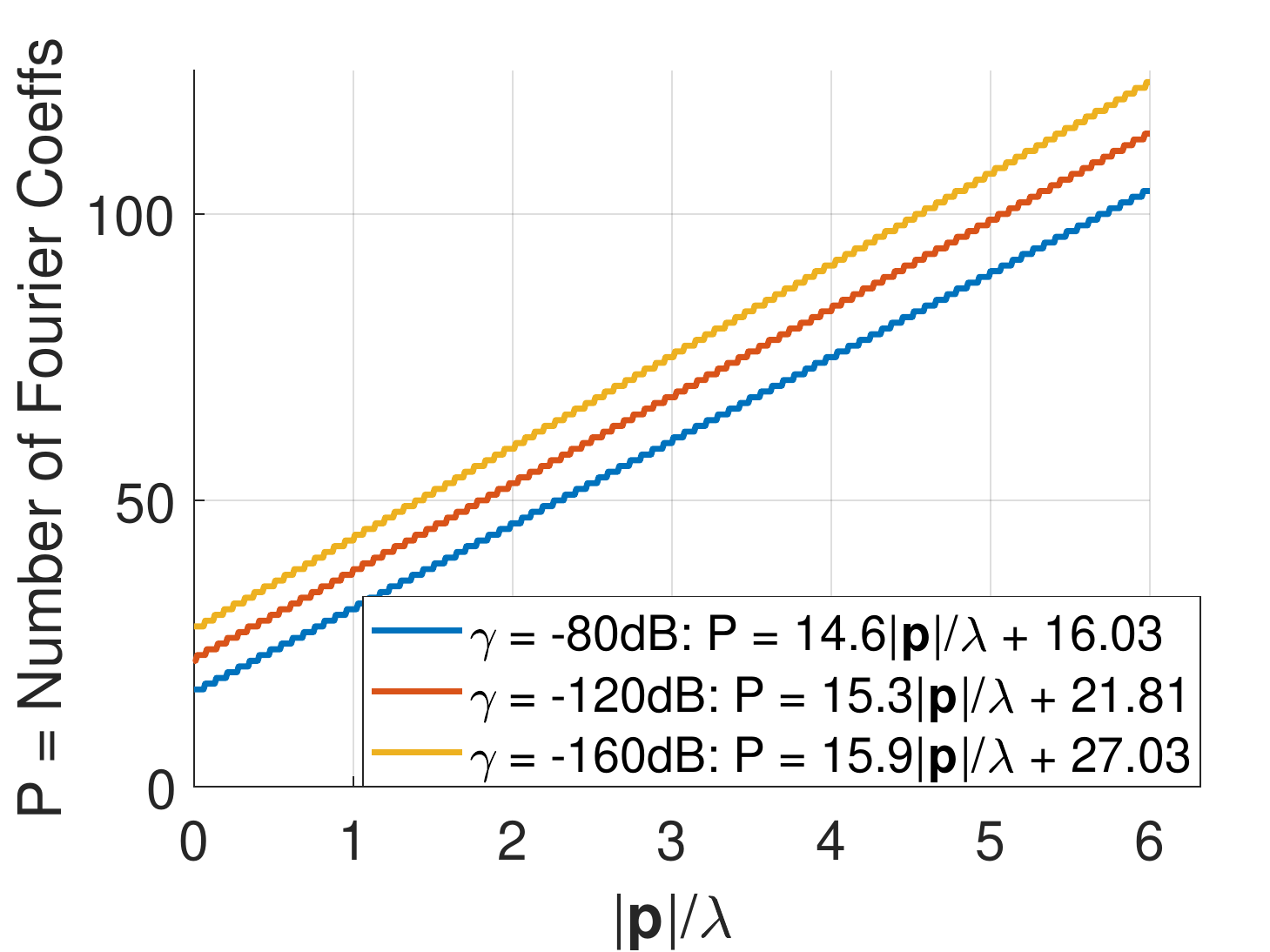}
			\put (17,63 ) {{\small {(\textbf{b})}}}
		\end{overpic}
	\end{subfigure}
	\vspace*{-1mm}
	\caption{ (a) Squared magnitude (dB) of FS coefficients as a function of $k$, the DFT index,  and $ |\bm{p}|/\lambda $, 
		(b) minimum $ P $ vs.~$ |\bm{p}|/\lambda $ for different FS magnitude cutoff levels ($\gamma$).}
	\label{fig:Fig1}
\end{figure}

In order to determine the value of $P$ needed for various array geometries, we conduct a numerical study of the FS based on the continuous function $a^*_m(\theta)$ defined in (\ref{eq:ar-spDefined},\,\ref{eq:ar-spExponent}). 
By using a very long DFT, the FS coefficients of 
$ a^*_m(\theta) $ can be computed numerically to get $ \hat\alpha[k] $.
From the exponent of $a^*_m(\theta)$ in \eqref{eq:ar-spExponent},
it is true that the magnitude $|\hat\alpha_m[k]|$ depends only on $ |\bm{p}|/\lambda $, because $(\theta-\angle\bm{p})$ is a shift in the argument of $a_m(\theta)$ which changes only the phase of its FS coefficients.
Thus, we use a long DFT to obtain FS coefficients for many different values of $ |\bm{p}|/\lambda $, and display the magnitude $|\hat\alpha[k]|^2$ as an image in  \cref{fig:Fig1}a, which confirms that $\hat\alpha[k]$ is bandlimited.
\vspace{-2mm}
A vertical cut through the image is the squared FS magnitude 
for one value of  $ |\bm{p}| / \lambda $, 
the normalized distance of the sensor from the reference point.
Along the horizontal axis  $ |\bm{p}| / \lambda $ is increasing, and the bandwidth of the FS grows.
Thus the distance of the farthest sensor from a reference point 
controls the minimum $ P $ needed to get an accurate DFT representation.
The index $N$ where $|\hat\alpha[k]| \approx 0$ for $|k|>N$  depends on choosing a threshold $ \gamma$ for 
the squared magnitude of the FS. 
\Cref{fig:Fig1}b shows three cases at $-80$, $-120$, or $-160$\,dB below the maximum.
For $ |\bm{p}| / \lambda \geq 2$, a linear approximation gives an excellent estimate for $P = 2N+1$.
This minimum value of $ P $ is important for reducing the computational complexity of the SDP.
For example, at $\gamma =-160$\,dB the disregarded FS coefficients are all below $10^{-8}\max{|\hat\alpha[k]|}$, and the linear estimate is $P = 15.9 |\bm{p}| / \lambda +27.03$.

Using the DFT representation in \eqref{eq:P-ptDFT}, the dual function $ b(\theta) $ can be related to a \textit{dual polynomial} $ \hat b(z) $ as
\begin{align}\label{eq:dualfunction_approx}
b(\theta)
&\simeq \sum_{k=-\NN}^{\NN}\hat{B}_{k}e^{jk\theta}
= \sum_{k=-\NN}^{\NN}\hat{B}_{k}z^{k} 
\;\buildrel{\Delta}\over{=}\;\hat b(z)\bigg{|}_{z=e^{j\theta}}
\end{align}
Combining \eqref{eq:BkFromAlphak} and \eqref{eq:dualfunction_approx}, we recognize that the coefficients $\hat{B}_{k}$ can be written in matrix-vector form with $\bm{h}\in\mathbb{C}^{P}$ being
\begin{equation}
\label{eq:Gc}
{\bm{h}} = \left[ \hat{B}_{-\NN} \;\;  \hat{B}_{-(\NN-1)} \;\ldots\;  \hat{B}_{\NN} \right]^T = \bm{G}^H \bm{c},
\end{equation}
where  $\bm{G}^H =  \begin{bmatrix} \hat\alpha_m[k] \end{bmatrix}_{P\times M}$ is a matrix  whose $m$-th column contains the FS coefficients of $a^*_m(\theta)$, and $\bm{c}$ is the dual vector.

\vspace*{-2mm}
\subsection{Semidefinite Programming}

Using the uniform boundedness of the function $\mathcal{S}(\theta)^H \bm{c}$ in \eqref{eq:gridfree_dual}, and hence that of its FD representation given by the dual polynomial $\hat b(z) $, 
{we convert the infinite number of constraints in the dual problem \eqref{eq:gridfree_dual} into finite-dimensional 
matrix constraints similar to 
\cite{candes_math_theory_super_resoln_2014,dumitrescu_positivetrigpoly_2007}, and obtain
the SDP,}
\begin{align*}\label{eq:FDgridfree_SDP}
\min_{\bm{c},\bm{H}} \operatorname{\Re}\{\bm{c}^H\bm{y}\},\ \ \ 
&{\rm s.t.\ } 
\begin{bmatrix} \bm{H}_{P \times P} &  \bm{G}^H_{P \times M}\bm{c}^{\phantom{.}}_{M \times 1} \\
\bm{c}^H \bm{G} & 1 \end{bmatrix} \succeq 0,
		\tag{$ 11$}
\\
\label{eq:FDgridfree_SDP2}
&\ \sum_{i=1}^{P-j}\bm{H}_{i,i+j} = \left\{ \begin{array}{cl}
1, &\mbox{ $j = 0$} \\
0 &\mbox{ $j = 1, \ldots , P-1$}.
\end{array}\right. 
\end{align*}
The matrix $ \bm{H}_{P \times P}$ is a positive semidefinite matrix satisfying the constraints in \eqref{eq:FDgridfree_SDP}. 
{The finite SDP in \eqref{eq:FDgridfree_SDP} has $ n=(P+1)^2/2 $ optimization variables, and can be efficiently solved in polynomial time using interior-point methods \cite{vandenberghe_boyd_sdp_1996}.
The observed time complexity was found to be much less than the worst case $ \mathcal{O}(n^3) $.}
The dual polynomial $\hat b(z)$ is the desired output after the SDP, so its coefficient vector is constructed from the optimal $ \bm{c}_* $ via 
$ \bm{h}_* = \bm{G}^H \bm{c}_* $. 
\vspace*{-2.3mm}
\subsection{DOA Estimation via Polynomial Rooting}
For sufficiently large $ P $, the representation of the dual function $ \mathcal{S}(\theta) ^H \bm{c}_*  $ by the dual polynomial $\hat b(e^{j\theta}) $ is highly accurate.
From the constraint \eqref{eq:gridfree_dual}, we have $\lvert \mathcal{S}(\theta) ^H \bm{c}_* \rvert = 1$ \text{for the true DOAs} $\theta \in [\theta_1,\ldots, \theta_L]$, and  $ \lvert \mathcal{S}(\theta) ^H \bm{c}_* \rvert <1 $ for other angles.
In other words, the magnitude of $\hat b(e^{j\theta}) $ is equal to one only for true DOAs, and less than one elsewhere.
Using this property, the DOAs are estimated by locating the angles $ \theta $ where the magnitude of the dual polynomial is one \cite{candes_math_theory_super_resoln_2014,xenaki_gridfree_2015}.
To accomplish this, we form a nonnegative polynomial 
\begin{equation} \label{eq:p_z}
p(z) = 1- |\hat b(z)|^2  
\tag{$ 12$}
\end{equation}
from the dual polynomial coefficients $ \bm{h}_* $.
The coefficients of $|\hat b(z)|^2$  are the autocorrelation coefficients of $ \bm{h}_* $, i.e.,  $ r_k  = \sum_j h_j h_{j-k}^*	$.
Finally, the angles of the zeros of $ p(z) $ on the unit circle are the DOAs of the sources.

The source amplitudes are recovered via least squares \cite{candes_math_theory_super_resoln_2014},
\begin{equation}\label{eq:est_amplitudes}
\hat{\bm{s}} =  \bm{A}(\hat{\bm{\theta}})^\dagger \bm{y},
\tag{$ 13$}
\end{equation}
where $ ^{\dagger} $ denotes the pseudo-inverse. The columns of the matrix $ \bm{A}(\hat{\bm{\theta}}) $ are the steering vectors for the estimated DOAs $ \hat{\bm{\theta}} $.

\Cref{alg1} lists the steps involved in the proposed method.
\begin{algorithm}
	\caption{\textbf{Super-resolution DOA for arbitrary array}}
	\SetKwInOut{KwInit}{Inputs}
	\AlFnt
	\KwIn{ Array snapshot vector $\bm{y} \in \mathbb{C}^M$, wavelength
	 $\lambda$, number of Fourier coefficients $ P $}
 \vspace*{-3mm}
	\hrulefill
	\begin{flushleft}
		1.\enskip {For the geometry, compute $ \bm{G}^H = \begin{bmatrix} \hat\alpha_m[k] \end{bmatrix}_{\tiny{P \times M}}$ via \eqref{eq:P-ptDFT}.}
		
	    2.\enskip {Using $ \bm{G}^H $ and $ \bm{y} $ as inputs, solve the SDP in \eqref{eq:FDgridfree_SDP} to \\ \quad \, find optimal $ \bm{c}_* $.} 
		
		3.\enskip {Compute the optimal dual polynomial coefficients-\\ \quad \, vector $ \bm{h}_*$, using  $ \bm{h}_*  = \bm{G}^H \bm{c}_*  $.  }
		
	    4.\enskip {Estimate DOAs $ \hat{\bm{\theta}} $ by finding the roots on the unit-  \\ \quad \, circle of the nonnegative polynomial $ p(z) $ 
		in \eqref{eq:p_z}.}		
	
		5.\enskip {Recover source amplitudes $ \hat{\bm{s}} $ using \eqref{eq:est_amplitudes}. }	
		 \vspace*{-2mm}
					
	\label{alg1}
	\end{flushleft}
\end{algorithm}

\vspace{-6mm}
\section{Simulations}

Results using \Cref{alg1} for the uniform circular array (UCA) and random planar array (RPA) geometries 
are presented in \Cref{subsec:UCA,subsec:RPA}. 
Performance is compared with the conventional delay-sum beamformer (CBF). 
\Cref{subsec:success_probability} presents two probability of success studies 
for various parameter values.
All simulations consider 
a single snapshot and multiple coherent sources \cite{shan_spatialsmoothing_1985}, which are complex sinusoids of the same frequency with constant phase difference.
We implemented the SDP \eqref{eq:FDgridfree_SDP} using CVX \cite{grant_boyd_cvx_2008}.
For DOA estimates, we use only the roots of $ p(z) $ that lie within a distance of $ 0.02 $ from the unit circle; the root angles are the DOAs.

\subsection{Simulations for Uniform Circular Array (UCA)}\label{subsec:UCA}
Two examples using a 40-element UCA are presented here. 
The array radius is $r= 2\lambda $, and 
the uniform sensor separation is $ d = (\pi/10)\lambda $.
With the reference point at the center of the array, $|\bm{p}_m| =r$ for all sensors.

\begin{figure}[htbp]
	\begin{subfigure}{.5\columnwidth}
		\centering
		\includegraphics[width=1\linewidth]{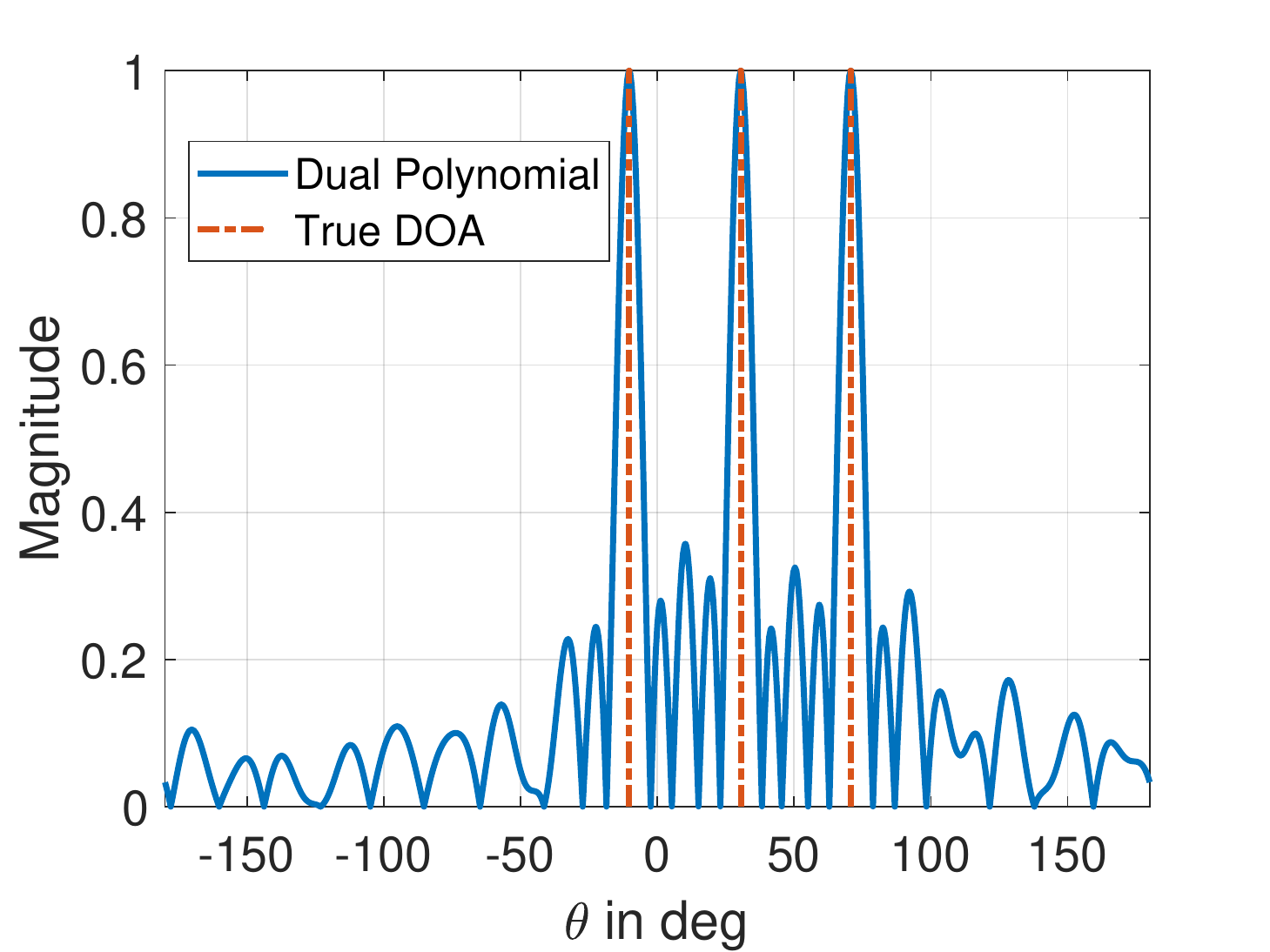}
		\caption{Dual polynomial}
		\label{fig:Fig2a}
	\end{subfigure}%
	\begin{subfigure}{.5\columnwidth}
		\centering
		\includegraphics[width=1\linewidth]{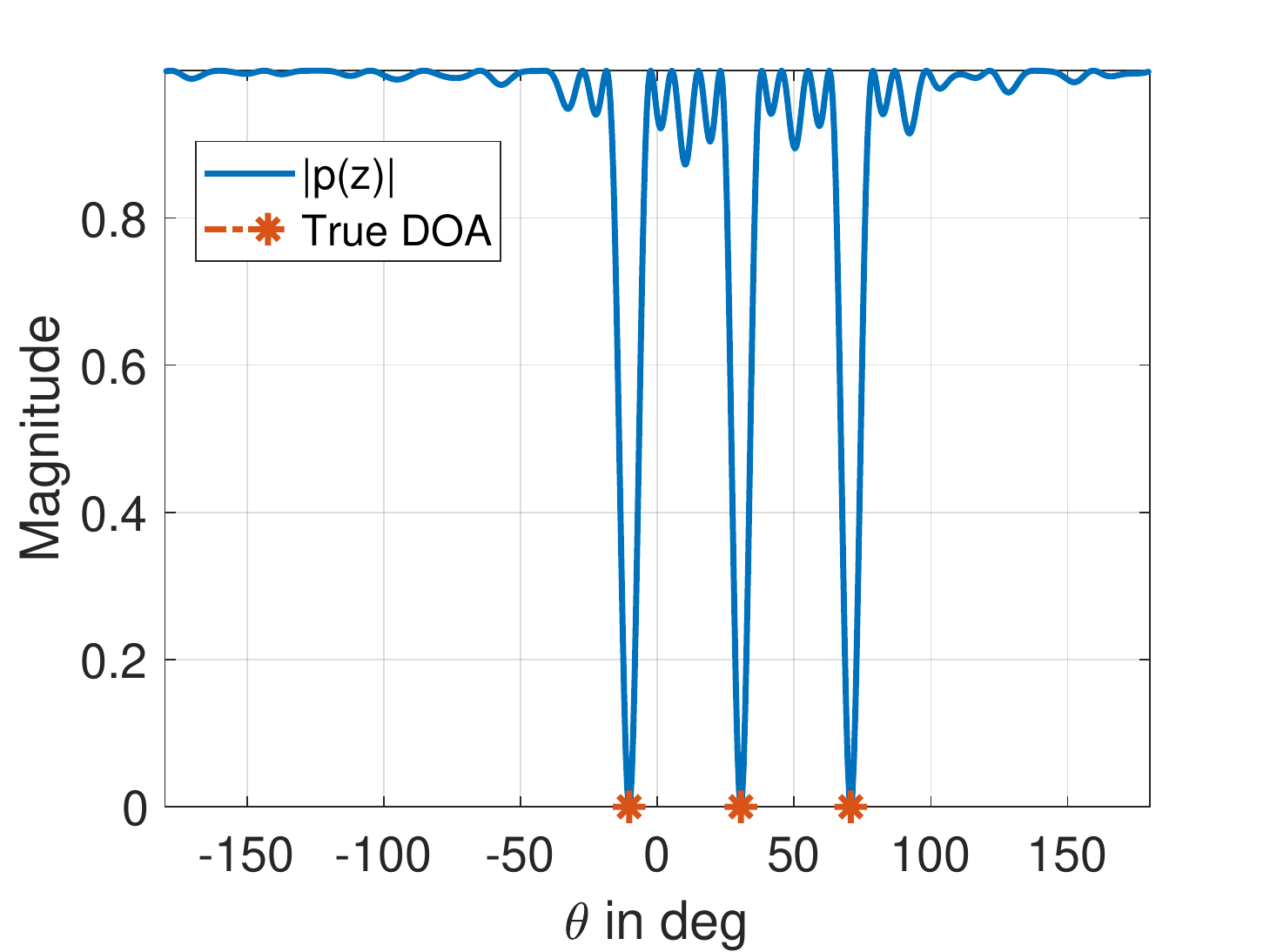}
		\caption{Nonnegative Polynomial}
		\label{fig:Fig2b}
	\end{subfigure}
	\\
	\begin{subfigure}{.5\columnwidth}
		\centering
		\includegraphics[width=1\linewidth]{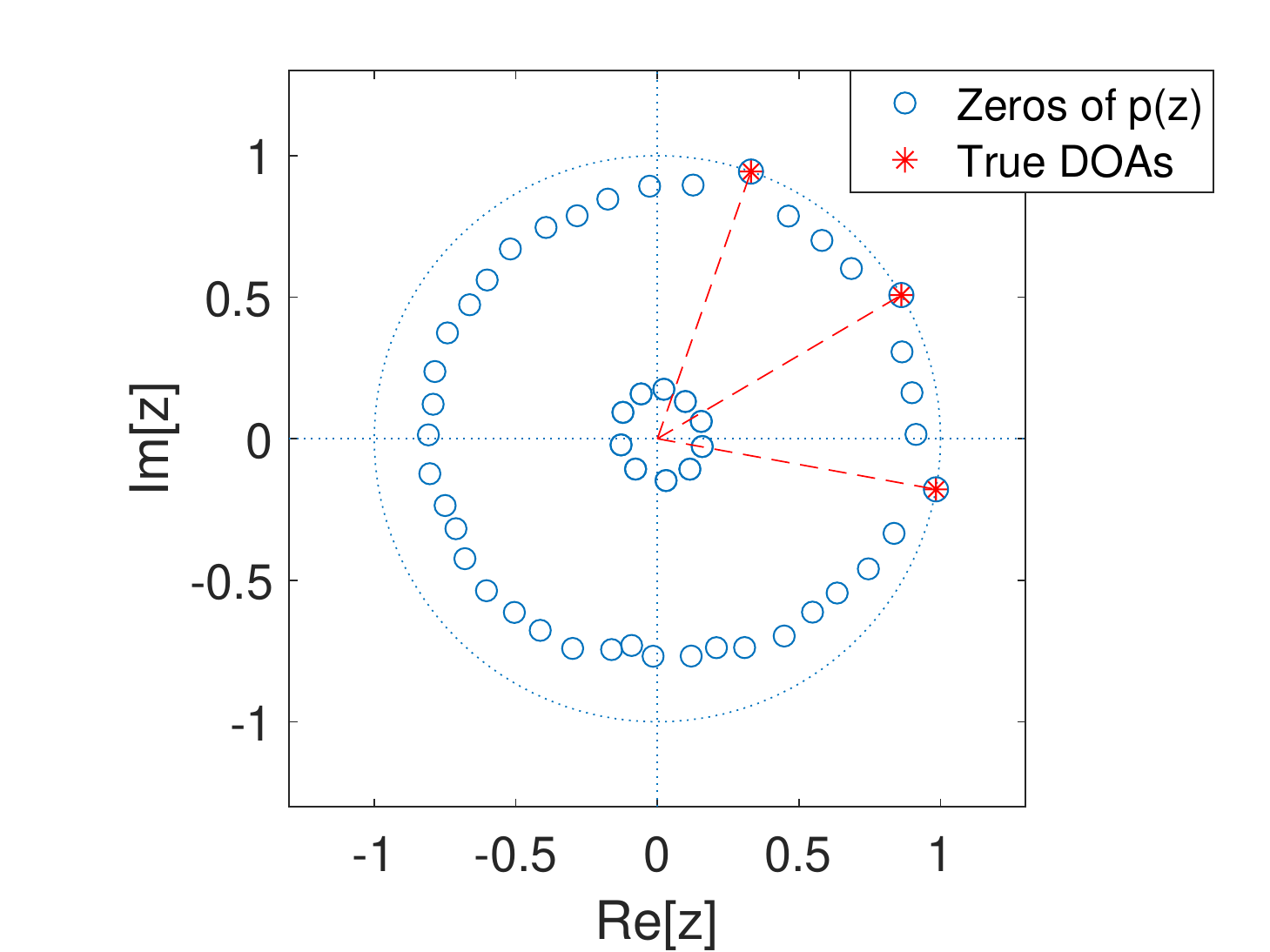}
		\caption{Zeros of $ p(z) $}
		\label{fig:Fig2c}
	\end{subfigure}%
	\begin{subfigure}{.5\columnwidth}
		\centering
		\includegraphics[width=1\linewidth]{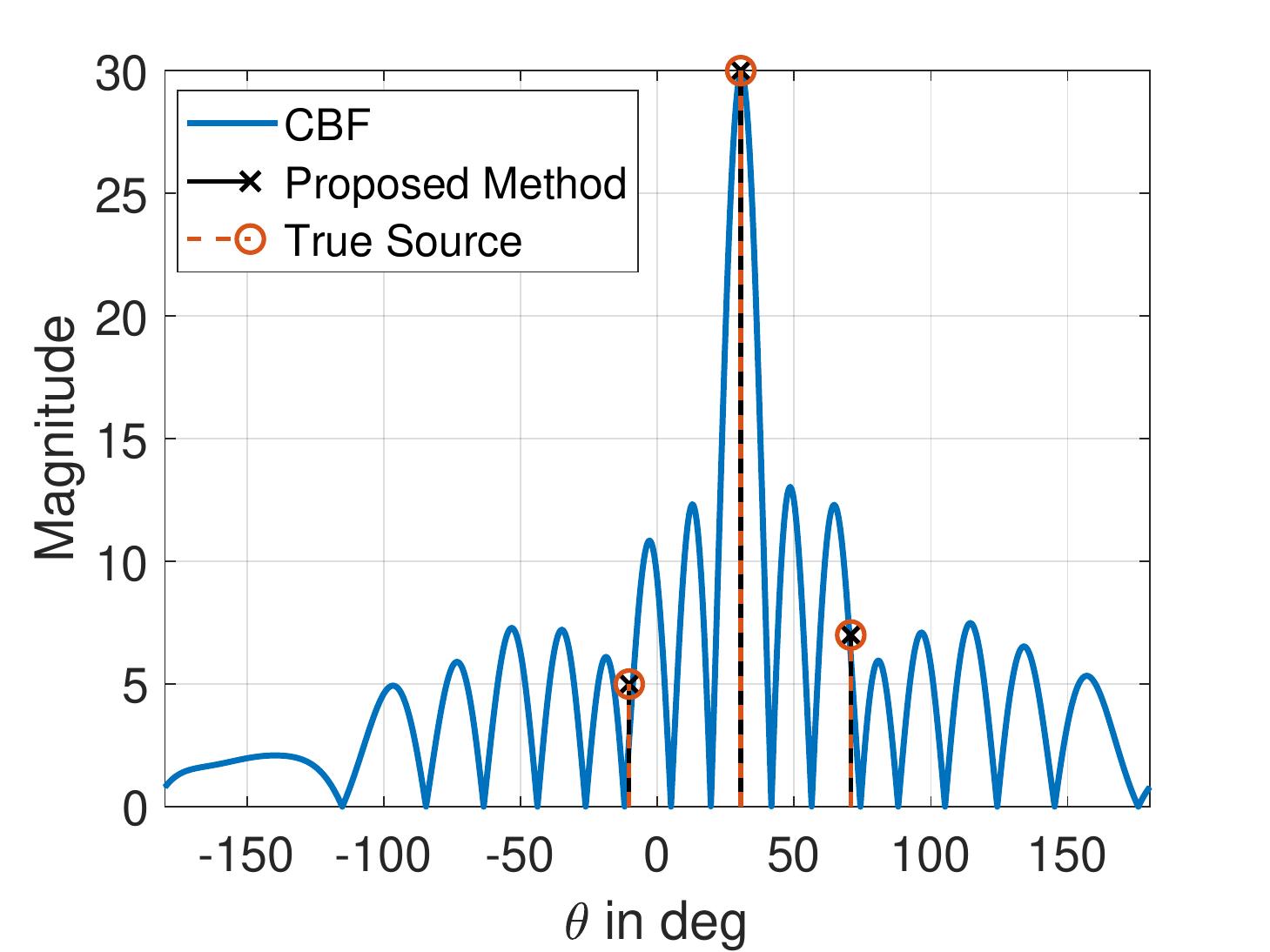}
		\caption{CBF vs. Proposed Method}
		\label{fig:Fig2d}
	\end{subfigure}
	
	\caption{Result for UCA with $M=40$, $P=61 $. Three sources at $ -10.3^\circ, 30.5^\circ, 70.7^\circ $ have magnitudes of $5, 30, 7$.
	In (c) only the zeros inside and on the unit circle are shown; the rest are at conjugate reciprocal locations $(1/z^*)$.
	} 
	\label{fig:Fig2}
\end{figure}

In the first example, we consider a case with three sources having different magnitudes.
The dual polynomial $\hat b(e^{j\theta}) $, and the nonnegative polynomial $ p(e^{j\theta}) $ are shown in \cref{fig:Fig2}a,b.
The DOAs are perfectly estimated using angles of the unit-circle zeros of $p(z)$ as shown in \cref{fig:Fig2c}.
Polynomial rooting eliminates searching over a fine grid of all angles.
The source amplitude estimates are also perfect.
As seen in \cref{fig:Fig2d}, the CBF is unable to estimate two of the three sources, due to the smaller magnitudes of those sources. 
The perfect DOA estimates of the proposed method validate its ability to estimate DOAs accurately for an arbitrary array geometry.
 
\begin{figure}[htbp]
	\begin{subfigure}{.5\columnwidth}
		\centering
		\includegraphics[width=1\linewidth]{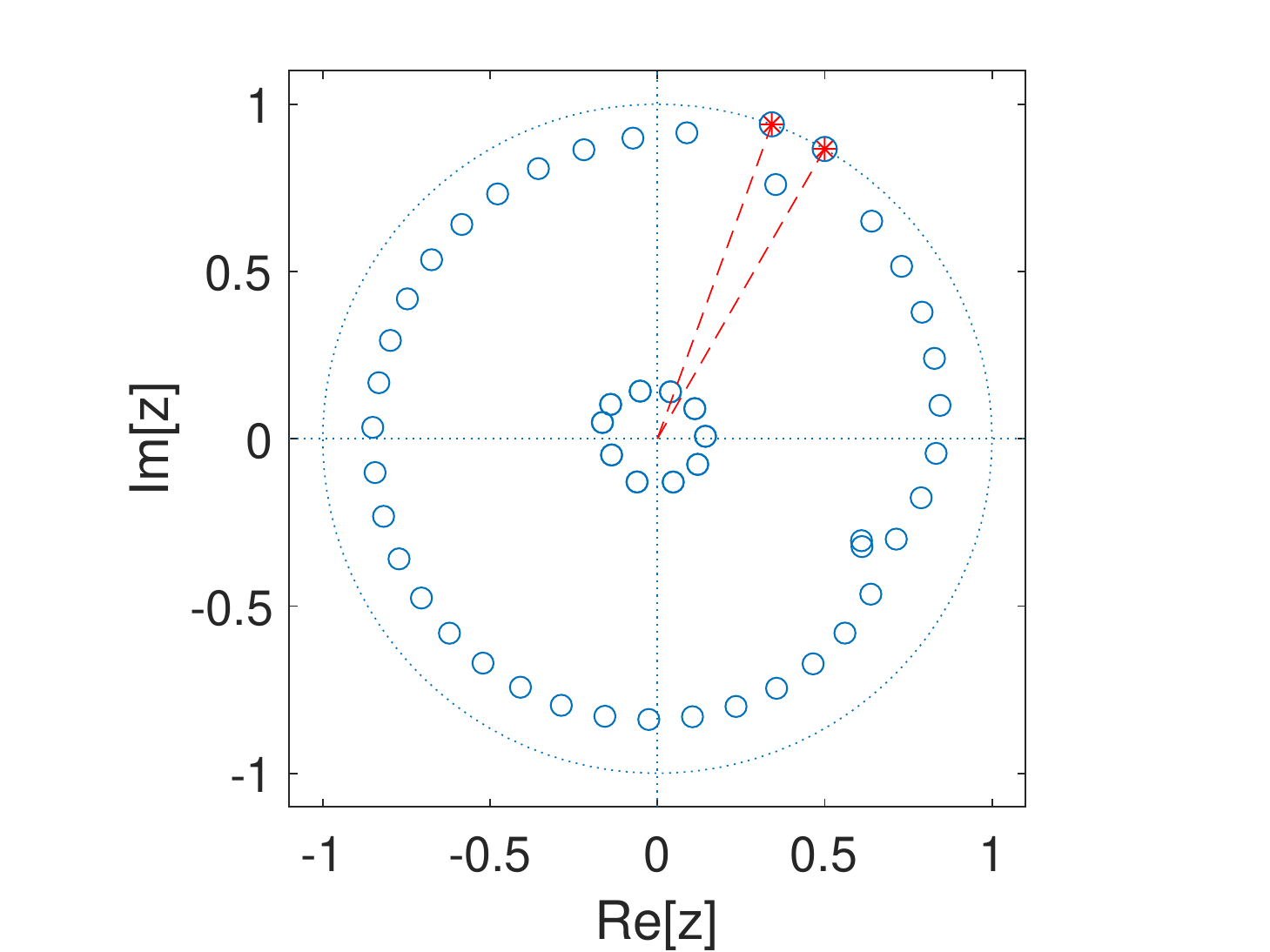}
		\caption{Zeros of $ p(z) $}
		\label{fig:Fig3a}
	\end{subfigure}%
	\begin{subfigure}{.5\columnwidth}
		\centering
		\includegraphics[width=1\linewidth]{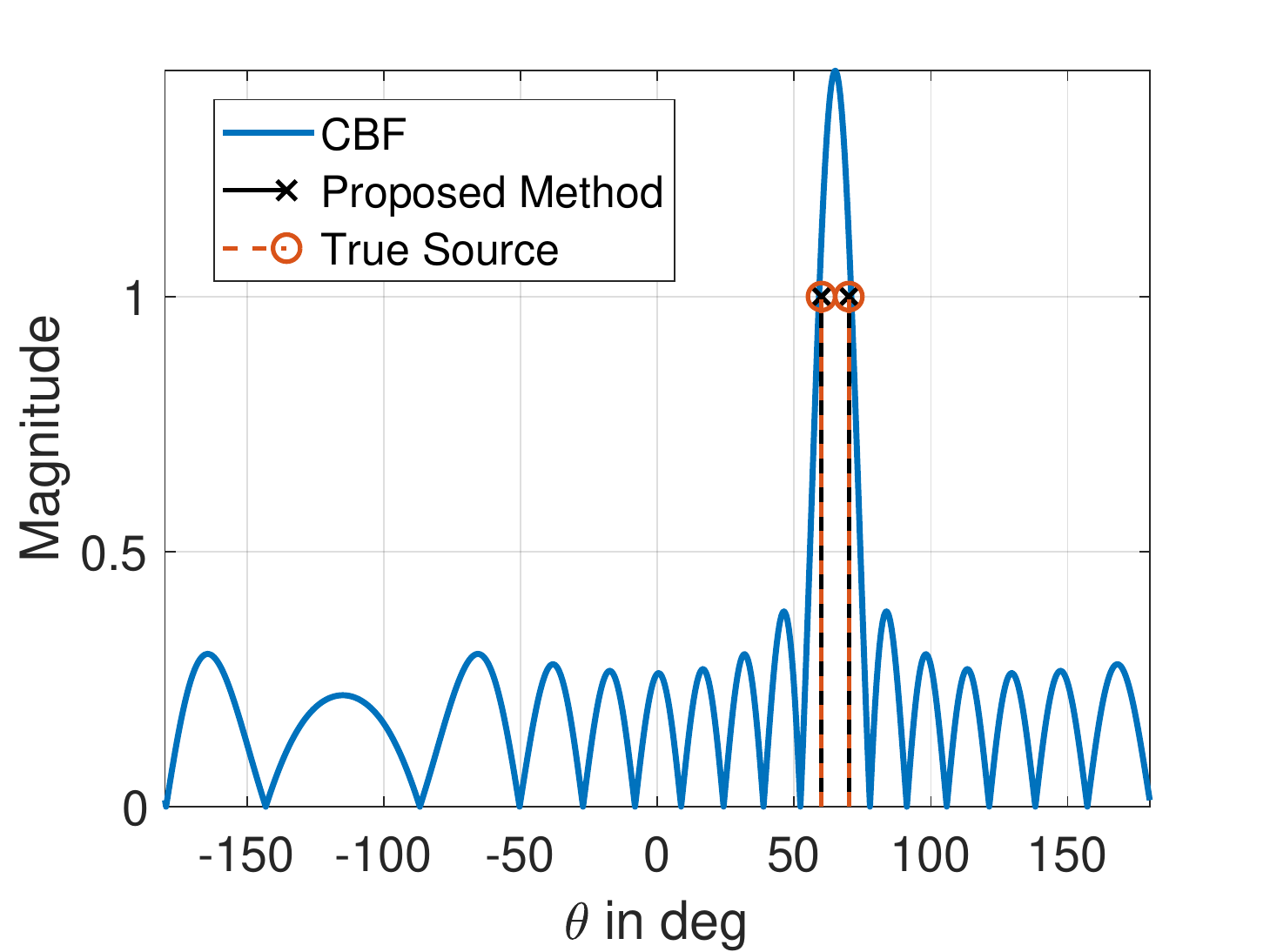}
		\caption{CBF vs. Proposed Method}
		\label{fig:Fig3b}
	\end{subfigure}
	
	\caption{Result for UCA with $M=40$, $P=61 $. Two equal magnitude sources with DOAs of $ 60^\circ $ and $ 70^\circ$. }
	\label{fig:Fig3}
\end{figure}

In \cref{fig:Fig3} we study the source resolution performance of the proposed approach considering two equal magnitude sources separated by $ 10^\circ $. 
As seen in \cref{fig:Fig3b}, the CBF is not able to resolve the two closely located sources, whereas estimates from the unit-circle zeros in \cref{fig:Fig3a} are perfect. 
This reinforces the fact that the proposed approach offers higher resolution than existing methods for single snapshot DOA estimation and justifies the term ``super-resolution.''

\vspace*{-6mm}
\subsection{Simulation for Random Planar Array (RPA)}\label{subsec:RPA}

\begin{figure}[htbp]
	\begin{subfigure}{.49\columnwidth}
		\centering
		\includegraphics[width=1\linewidth]{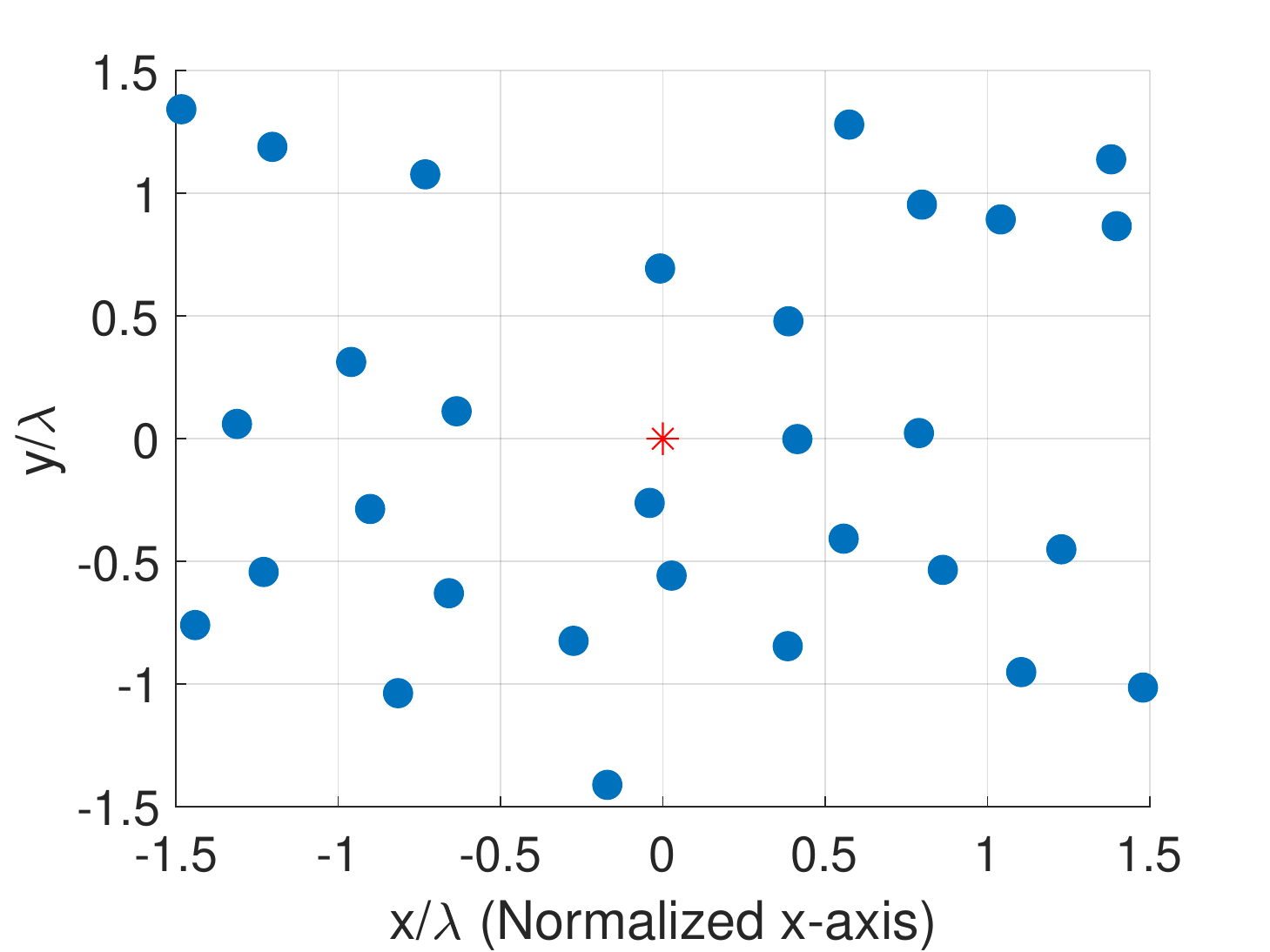}
		\caption{Random Planar Array (RPA)}
		\label{fig:Fig4a}
	\end{subfigure}
	\begin{subfigure}{.49\columnwidth}
		\centering
		\includegraphics[width=1\linewidth]{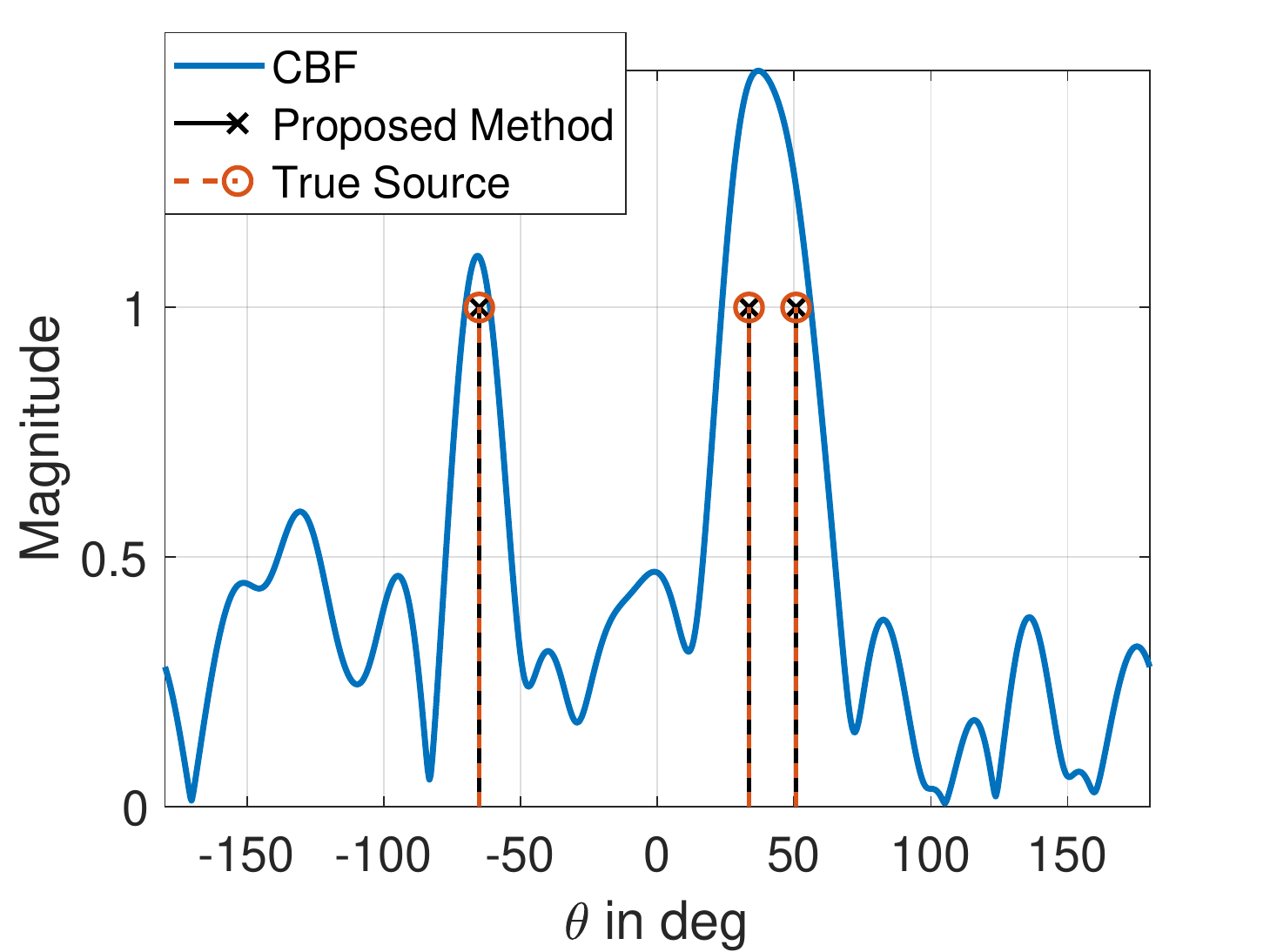}
		\caption{CBF vs. Proposed Method}			
		\label{fig:Fig4b}
	\end{subfigure}
	\caption{Result for RPA with $ M=30$, $P=61 $, and three equal magnitude  sources with DOAs at $ -65.1^\circ$, $37.5^\circ$, and $50.7^\circ $. }
	\label{fig:Fig4}
\end{figure}

In \cref{fig:Fig4a}, we consider a random planar array with  $ 30 $ sensors.
The minimum sensor spacing is $ d = \lambda/4 $, and the distance of the farthest sensor from the origin is $ 2 \lambda $.
The CBF is unable to resolve two of the sources and the amplitudes are inaccurate,
whereas the proposed method perfectly estimates all the sources.
This example verifies that the proposed method is applicable to any arbitrary geometry.

\vspace*{-2mm}

\subsection{Performance Evaluation using Success Probability}    \label{subsec:success_probability}

\begin{figure}[htbp]
	\begin{subfigure}{.49\columnwidth}
		\centering
\includegraphics[width=1\linewidth]{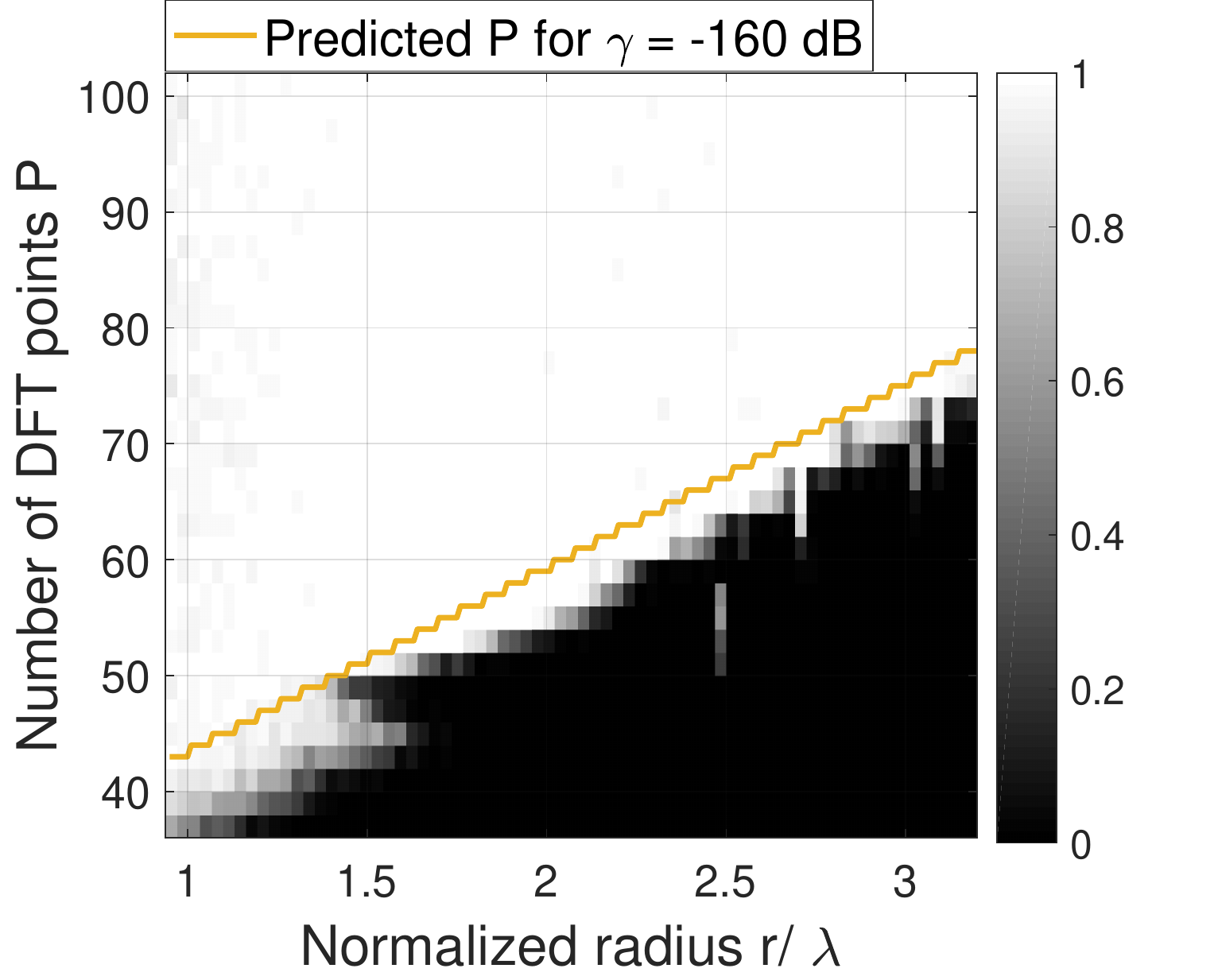} 
		\caption{}
		\label{fig:Fig5a}
	\end{subfigure}
	\begin{subfigure}{.49\columnwidth}
		\centering
			\includegraphics[width=1\linewidth]{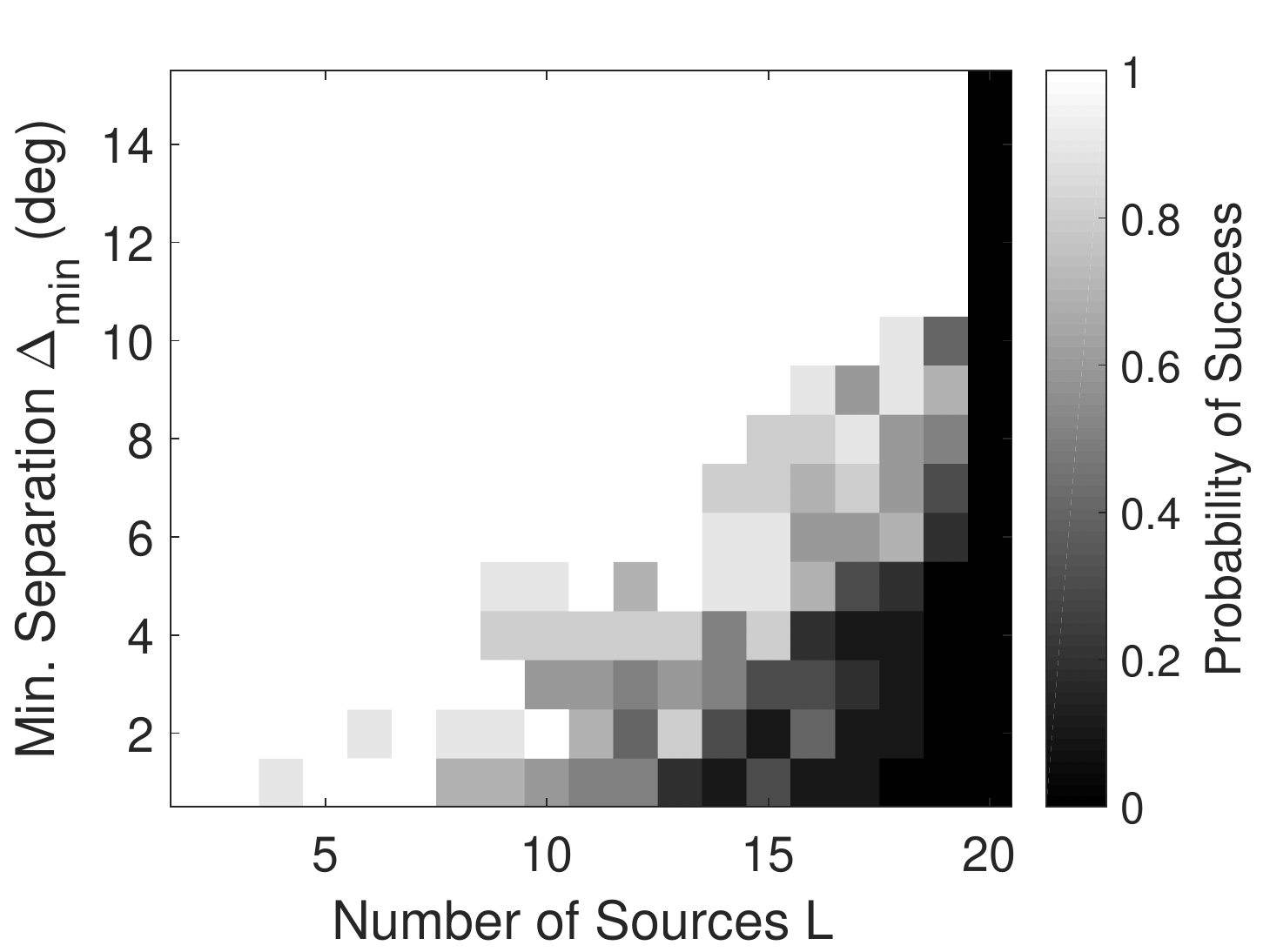} 
		\caption{}
		\label{fig:Fig5b} 
	\end{subfigure}
	\caption{ Success probability of $ M=40$ UCA (a) versus $r/\lambda$ and $ P $, (b) versus minimum source separation $ \Delta_{\min} $ and $ L $.  }
	\label{fig:Fig5}
\end{figure}

We now study the DOA estimation performance of the overall approach by measuring ``success'' in terms of exact recovery for various values of $ P $ and UCA apertures (radii).
\Cref{fig:Fig5a} shows the probability of successfully recovering the DOAs of $ 10 $ sources with equal complex amplitudes,
using a $ 40 $-sensor UCA.
The source DOAs are generated from a uniform distribution in $(-\pi,\pi]$, such that the minimum 
\textit{wraparound}\footnote{The wraparound separation between sources at $ -175 ^\circ $ and $ + 177 ^\circ $ is $ 8^\circ $. }
source separation $ \Delta_{\min} $ is $ 10^{\circ} $. 	
Fifty random trials are run for each $  P $ and $ r/\lambda $.
Success is declared when DOAs of all the sources are estimated within $ {0.001}^\circ $ error, otherwise, a trial is labeled as failure.
The probability of success is shown as an image in \cref{fig:Fig5a}.
We also show an overlay plot of the predicted $ P $ for $\gamma = -160 $\,dB from \cref{fig:Fig1}b,
which validates the accuracy of the prediction and provides additional evidence that the minimum $ P $ is linearly related to $ r $.

\Cref{fig:Fig5b} studies the success rate versus the number of sources $  L $ and $ \Delta_{\min} $ for a 40-sensor UCA with a fixed radius of $ r/\lambda  = 1.59 $  ($ d=\lambda/4 $), using the same success criterion, and $ 10 $ trials.
$ P $ is set equal to 
$ 53 $ for $ r/\lambda  = 1.59 $, based on $ \gamma = -160 $\,dB in \cref{fig:Fig1}b.
In \cref{fig:Fig5b}, as $ L $ increases, a larger $ \Delta_{\min} $ is required for success.
However, for $ \Delta_{\min} > 10^{\circ}$ and $ L < {M}/{2} $, we have exact DOA recovery of all sources.
This study confirms the existence of a minimum source separation condition \cite{candes_math_theory_super_resoln_2014} and shows the limit on the maximum number of sources \cite{fuchs_sparsity_2005} for exact recovery.

\vspace*{-2.3mm}
\section{Discussion}
We have described a super-resolution DOA estimation 
method for arbitrary geometry arrays, which is applicable for single snapshots and correlated or uncorrelated sources.
DOA estimates are obtained from unit circle zeros of a nonnegative polynomial formed from a dual polynomial.
The periodicity of the array manifold vs.\ angle $\theta$ leads to a finite Fourier series representation and a finite degree dual polynomial.
The number of FS coefficients required depends primarily on the distance to the sensor farthest from a reference.
{Thus it would be important to choose the reference point at (or near) the array center to reduce computational complexity, which increases as the array size increases.
Furthermore, the FD method would not be used for ULAs because existing sparsity based gridless super-resolution approaches directly provide a trigonometric polynomial for the dual function. }

We have treated only the noise-free case here, but the same approach applies when the model in \eqref{eq:MeasurementModel} includes an additive noise vector.
The dual problem has the same constraint as in \eqref{eq:dualfunction_approx}, so the FS representation is identical, depending only on the distance from a reference.
Simulations with a modified SDP using the approach of \cite{candes_super_resoln_noisy_2013} for noisy observations and multiple snapshots will be reported in a longer paper.

{We only compared the proposed method with the CBF, because the high resolution adaptive DOA approaches \cite{capon_mvdr_1969,schmidt_music_1986,stoica_spice_likes_2012,ottersten_stoica_covariancematching_1998} fail in single snapshot and\,/\,or coherent signal conditions, though they are applicable for arbitrary arrays.}
Simulation results prove the applicability of the proposed method for high resolution search-free DOA estimation for arbitrary geometries, 
using a single snapshot.
We considered estimation of azimuth DOAs here. It may be possible to form a 2D polynomial to estimate both azimuth and elevation angles.

\FloatBarrier

\bibliographystyle{IEEEtran}
\bibliography{Refs}
\end{document}